\documentclass[prd,preprint,nofootinbib,prd]{revtex4}

\usepackage{epsf,epsfig,subfigure,graphicx,amsmath,amssymb}
\usepackage{amsfonts}
\usepackage{latexsym}
\usepackage{color}
\usepackage{accents}

\def\bea{\begin{eqnarray}}
	\def\eea{\end{eqnarray}}

\def\Mp{M_{\rm Pl}}

\begin{document}
\title{Aligned natural inflation with modulations}
\author{
Kiwoon Choi$\,^a$\footnote{kchoi@ibs.re.kr} and 
Hyungjin Kim$\,^{b,a}$\footnote{hjkim06@kaist.ac.kr}
}
\affiliation{
$^a$ Center for Theoretical Physics of the Universe,  Institute for Basic Science (IBS), Daejeon, 34051, Korea\\
$^b$ Department of Physics, KAIST, Daejeon, 305-701, Korea 
}
\begin{abstract}
The weak gravity conjecture applied for the aligned natural inflation indicates that generically there can be a  modulation of the inflaton potential, with a period determined by sub-Planckian axion scale. We study the oscillations in the primordial power spectrum induced by such modulation, and discuss the resulting observational constraints on the model.
\end{abstract}
\preprint{CTPU-15-19}
\maketitle

\section{Introduction}\label{intro}
Inflation in the early universe explains the flatness, horizon, and entropy problems in the standard big bang cosmology, while providing a seed of the large scale structure and the anisotropy in cosmic microwave background (CMB) radiation observed today. If the energy scale of inflation is high enough,  the de Sitter quantum fluctuation of spacetime metric can give rise to a primordial tensor perturbation which might be large enough to be detectable in the near future. On the other hand, such high scale inflation  demands a super-Planckian excursion of the inflaton~\cite{LythBound}, so a scalar potential which is flat over a super-Planckian range of the inflaton field. In view of the UV sensitivity of scalar potential, this is a nontrivial condition required for the underlying theory of inflation.

As is well known,  a pseudo-Nambu-Goldstone boson $\phi$  can have a naturally flat potential over a field range comparable to its decay constant $f$. The low energy potential is protected from unknown UV physics under the simple assumption that  UV physics respects  an approximate global symmetry which is non-linearly realized in the low energy limit as $\phi/f \rightarrow \phi/f+c$, where $c$ is a real constant. In the natural inflation scenario~\cite{Freese:1990rb,Adams:1992bn},  inflaton is assumed to be a pseudo-Nambu-Goldstone boson having a sinusoidal potential generated by non-perturbative dynamics. Then the inflationary slow-roll parameters have a size of  ${\cal O}(\Mp^2/f^2)$, and therefore  the model requires  $f\gg  \Mp$, where $\Mp\simeq 2.4\times 10^{18}$ GeV is the reduced Planck scale. Although it appears to be technically natural within the framework of effective field theory, there has been a  concern that the required super-Planckian decay constant may not have a UV completion consistent with quantum gravity~\cite{ArkaniHamed:2006dz}. Also, previous studies on the axion decay constants in string theory suggest that generically  $f<\Mp$, at least in the perturbative regime~\cite{Choi:1985je,Banks:2003sx,Svrcek:2006yi}.

Nevertheless, one can engineer the model  to get a super-Plackian axion decay constant within the framework of effective field theory \cite{Kim:2004rp,Dimopoulos:2005ac, Kaloper:2008fb,Kaloper:2011jz}, or even 
in string theory \cite{Silverstein:2008sg,McAllister:2008hb,Andriot:2015aza}.
  An interesting approach along this direction is the aligned natural inflation {\cite{Kim:2004rp,Harigaya:2014eta,Choi:2014rja,Higaki:2014pja,Tye:2014tja,Kappl:2014lra,Ben-Dayan:2014zsa,Long:2014dta,Higaki:2014mwa, Shiu:2015uva,Gao:2014uha}}. In this scheme, initially one starts  with multiple axions, all having a sub-Planckian decay constant. Provided that the axion couplings  are  aligned to get a specific form of  axion potential \cite{Kim:2004rp}, a helical flat direction with multiple windings is formed in the multi-dimensional field space with  sub-Planckian volume.  If the number of windings is large enough, this flat direction can have a super-Planckian length, so result in an inflaton with super-Planckian effective decay constant.

Recently there has been a renewed interest in  the implication of the weak gravity conjecture (WGC)~\cite{Rudelius:2015xta,Montero:2015ofa,Brown:2015iha,Heidenreich:2015wga,Junghans:2015hba} for the aligned natural inflation. The WGC was  proposed initially for $U(1)$ gauge interaction~\cite{ArkaniHamed:2006dz}, implying  that there should exist a charged particle  with mass $m$ and charge $q$ satisfying $q/m\geq 1/\Mp$, so the gravity should be weaker than the $U(1)$ gauge force\footnote{There are two different versions of the WGC, the strong and the mild.  The strong WGC requires that the mass and charge of the lightest charged particle satisfy $q/m\geq 1/\Mp$, while in the mild version  the required particle with $q/m\geq 1/\Mp$ does not have to be the lightest charged particle. Here we are mostly concerned with the mild version generalized to the case of multiple axions.}. When translated  to axions, the WGC suggests  that there should exist an instanton which couples to the corresponding axion with a strength stronger than the gravity. This leads to an upper bound on the decay constant of individual axion, which is given by $f\leq {\Mp}/S_{\rm ins}$, where $S_{\rm ins}$ is the Euclidean action of the instanton~\cite{ArkaniHamed:2006dz}.

It has been noticed \cite{Cheung:2014vva} that  for models with multiple $U(1)$ gauge interactions, the WGC often leads to a stronger constraint on the charge-to-mass ratios than the one obtained by considering the individual $U(1)$  separately. The reason is that the WGC  applies  for all directions in the multi-dimensional charge space, not just for the charge vectors of the individual particles. For the case of  multiple axions, one similarly obtains a stronger constraint on the axion couplings~\cite{Brown:2015iha}. One then finds~\cite{Rudelius:2015xta,Montero:2015ofa,Brown:2015iha,Heidenreich:2015wga} that the alignment mechanism {\it cannot} be compatible with the constraint from the WGC   {\it if} the axion-instanton couplings required by the WGC  coincide with  the couplings generating  the axion potential that implements the alignment mechanism. 

A simple solution to this problem is that some of the instantons required by the WGC  do not participate in implementing the alignment mechanism~\cite{Brown:2015iha}.  Since some part of the axion potential for the alignment mechanism could be induced by either perturbative effects such as flux or other forms of nonperturbative effects such as  hidden gaugino condensation, this appears to be a rather plausible possibility. Indeed, known string theoretic construction of the aligned natural inflation  involves an instanton  which is not relevant for the alignment mechanism {\cite{delaFuente:2014aca,Abe:2014xja,Hebecker:2015rya,Ruehle:2015afa,Kappl:2015esy,Palti:2015xra}}. Yet, such instanton can generate an additional axion potential which does not affect the alignment mechanism as it corresponds to a subleading correction, but may give rise to an observable consequence in the precision CMB data.  As we will see, in the presence of such instanton, the low energy inflaton potential is generically given by\footnote{{The so-called multi-natural inflation scenario \cite{Czerny:2014wza,Czerny:2014wua} assumes the same form of inflaton potential, but with $f_{\rm mod}\gtrsim \Mp/ 2\pi$ and $\Lambda_{\rm eff}\gtrsim \Lambda_{\rm mod}$, whose observational consequences are different from our case with $f_{\rm mod} < \Mp/2\pi$ and $\Lambda_{\rm eff}\gg \Lambda_{\rm mod}$.}}
\bea
V_{\rm eff}(\phi) = \Lambda_{\rm eff}^4 \left[ 1 - \cos \left(\frac{\phi}{f_{\rm eff}} \right) \right] + \Lambda_{\rm mod}^{4} \left[ 1 - \cos \left( \frac{\phi}{f_{\rm mod}} + \delta\right) \right],
\eea
where the first term with $f_{\rm eff}\gg \Mp$ corresponds to the dominant inflaton potential generated by a dynamics implementing the alignment mechanism, while the second term with  $f_{\rm mod} \ll \Mp$ and $\Lambda_{\rm mod}^{4} \ll \Lambda_{\rm eff}^4$  is a subleading modulation generated by an instanton which is required by the WGC, but does not participate in implementing the alignment mechanism.

In this paper, we first provide an argument implying that the presence of subleading modulation in the inflaton potential  is a generic  feature of the aligned natural inflation consistent with the WGC. We then study the observable  consequences of such modulation, while focusing on the parameter region favored by theoretical or phenomenological considerations.  Specifically we consider the region $f_{\rm eff} \gtrsim 5 \Mp$ to avoid a fine tuning of the initial condition, while being consistent with the CMB  data~\cite{Ade:2015lrj}, and $f_{\rm mod} \lesssim \frac{\Mp}{2\pi}$ which is suggested by the WGC.  Observable  consequences of modulation in the axion monodromy inflation were studied extensively in~\cite{Flauger:2009ab,Flauger:2014ana,Flauger:2010ja,Peiris:2013opa,Price:2015xwa}, where it was  noticed that modulation can lead to an oscillatory behavior of  the power spectrum of the primordial curvature perturbation. We examine the constraint from CMB on modulations for the case of aligned natural inflation, and find that the CMB data restrict the amplitude of modulation as $\Lambda_{\rm mod}^{4}/\Lambda_{\rm eff}^4\,\lesssim\, {\cal O}(10^{-4}-10^{-6})$, depending upon the value of $f_{\rm mod}$.

It has been pointed out that modulation may  significantly change the predicted value of  the tensor-to-scalar ratio $r$ in natural inflation scenario \cite{Abe:2014xja,Kappl:2015esy}. We find that the change of the predicted value of $r$ due to modulation is minor, e.g. at most of ${\cal O}(10)$\%, if the amplitude of modulation is within the range compatible with the observed CMB data\footnote{For an alternative scenario which can give rise to a significantly smaller $r$ within the (aligned) natural inflation, see \cite{Albrecht:2014sea,Peloso:2015dsa,Achucarro:2015caa}.}. On the other hand, including an oscillatory part of the curvature power spectrum in  data-fitting analysis, we find that a larger parameter region in the $(n_s,r)$ plane can be compatible with the  CMB data compared to the case without an oscillatory piece, where $n_s$ denotes the spectral index of the curvature power spectrum. This makes it possible that the potential tension between the CMB data and  the natural inflation scenario is ameliorated under the assumption that there exists a modulation of the inflaton potential yielding a proper size of oscillatory piece in the curvature power spectrum.

This paper is organized as follows. In section~\ref{model}, we revisit the weak gravity conjecture applied for models with multiple axions, as well as the Kim-Nilles-Peloso (KNP) alignment mechanism. We argue that a small modulation of the inflaton potential is a  generic feature of the aligned natural inflation compatible with the WGC.  In section~\ref{cmb}, we study the oscillations in the curvature power spectrum induced by modulation, and discuss the constraints on the model from the CMB data. Section~\ref{conclusion} is the conclusion.

\section{weak gravity conjecture and the KNP alignment}\label{model}

In this section, we revisit the weak gravity conjecture applied for the aligned natural inflation, as well as the  alignment mechanism. As we will see, the WGC implies that generically there can be a small modulation of the inflation potential, with a period determined by sub-Planckian axion scale.

Let us begin with the constraint on the axion couplings for models with multiple axions, which is referred to the {\it convex hull condition} (CHC)~\cite{Cheung:2014vva,Brown:2015iha}. It requires first that in the presence of $N$ axions,
$$\vec\phi=(\phi_1,\phi_2, ...,\phi_N),$$
there exist corresponding (at least) $N$ instantons generating axion-dependent physical amplitudes as
\bea
{\cal A}_I \,\propto \, \exp\left(-S_I +i\vec{q}_I\cdot \vec\phi\right) \quad (I=1,2,...,N),\eea
where $S_I$ denotes the Euclidean action of the $I$-th instanton, and the axion-instanton couplings $\vec{q}_I$ are linearly independent from each other. It is always possible to parametrize the axion-instanton couplings as 
\bea
\vec{q}_I\,=\, \left(\frac{n_{I1}}{f_1},\frac{n_{I2}}{f_2}, ..., \frac{n_{IN}}{f_N}\right),\eea
where $f_i$ ($i=1,2,...,N$) can be identified  as the decay constant of the $i$-th axion,  and $n_{Ii}$ are integer-valued model parameters, so  the instanton amplitudes are periodic under the axion shift:
\bea
\phi_i \, \rightarrow \, \phi_i + 2\pi f_i.\eea
In the following, we will assume for simplicity that all instanton actions have a common value bigger than the unity\footnote{Usually $S_I\propto 1/g^2$ for a certain coupling constant $g$, and then a strong-weak coupling duality on $g$ may provide a firm basis on our assumption that $S_{\rm ins}>1$.}, e.g.
$$S_I \,=\, S_{\rm ins}\,>\,1.$$ 
By taking an analogy to the case of multiple $U(1)$ gauge fields, it has been argued that the axion-instanton couplings should be stronger than the gravity in {\it all} directions in the $N$-dimensional coupling space. Specifically,  one finds that the convex hull spanned by
\bea
\vec{z}_I \,\equiv \, \frac{\Mp}{S_{\rm ins}}\vec{q}_I \eea
should contain the $N$-dimensional unit ball with a center at the origin. Equivalently, one needs
\bea
|\vec q_I| \, > \frac{S_{\rm ins}}{\Mp} \quad \mbox{for all}\,\,\, I,\eea
and  for an {\it arbitrary} unit vector $\vec u$, 
\bea
|\vec u\cdot \vec q_I|\, >\, \frac{S_{\rm ins}}{\Mp} \quad \mbox{for some}\,\,\, I.
\label{CHC}\eea

The above convex hull condition has an immediate consequence on the aligned natural inflation. To see this, let us consider a simple two axion model for the alignment mechanism, which has the following KNP-type  axion potential:
\bea
V_0 = \Lambda_1^4 \left[  1 - \cos \left( \vec{p}_1\cdot \vec \phi  \right) \right]
+ \Lambda_2^4 \left[ 1 - \cos \left( \vec{p}_2\cdot \vec \phi \right) \right], \eea
where the axion couplings  $\vec p_I$ ($I=1,2$)  can be parametrized as
\bea
 \vec{p}_1 =  \left( \frac{\tilde n_{11}}{f_1},\, \frac{\tilde n_{12}}{f_2} \right)\quad \mbox{and}\quad
\vec{p}_2 =  \left( \frac{\tilde n_{21}}{f_1},\, \frac{\tilde n_{22}}{f_2} \right)
\eea
with integer-valued  $\tilde n_{ij}$. For $\Lambda_1^4 \sim \Lambda_2^4$, which will be assumed in the following discussion, the above axion potential has an approximately flat direction in the limit that $\vec p_1$ and $\vec p_2$ are aligned to be nearly parallel:
\bea
\sin\theta_p\,=\,\frac{1}{|\vec p_1||\vec p_2|} \det \left(
\begin{array}{c}
\vec{p_1} \\
\vec{p}_2 
\end{array}\right) \, \ll \, 1.
\label{align}
\eea
A particularly convenient parametrization of this flat direction is provided  by 
\bea
{\phi}_{\rm inf} = \frac{1}{|\vec p_1 -\vec p_2|} \, \det \left(
\begin{array}{c}
\vec{\phi} \\
\vec{p}_1- \vec{p}_2 
\end{array}
\right)\,\equiv \, \vec \xi\cdot \vec \phi,
\eea
where the flat direction unit vector $\vec \xi$ is chosen to be orthogonal to $\vec p_1-\vec p_2$. After integrating out the heavy axion,  we are left with a light inflaton
\bea
\vec \phi = \phi_{\rm inf}\, \vec\xi\eea
with an effective potential 
\bea
V_0 = (\Lambda_1^4+\Lambda_2^4) \left[ 1-\cos\left(\frac{\phi_{\rm inf}}{f_{\rm eff}}\right)\right],\eea
where
\bea
f_{\rm eff}\, =\, \frac{1}{|\vec \xi\cdot \vec p_1|}\,=\, \frac{1}{|\vec \xi\cdot \vec p_2|}\,=\, \frac{|\vec{p}_1 - \vec{p}_2|}{\det \left( \vec p_1, \vec p_2\right)^T}.
\label{KNP}
\eea
If $\vec p_1$ and $\vec p_2$ are aligned to be nearly parallel, then the inflaton direction $\vec \xi$ becomes nearly orthogonal to both $\vec p_1$ and $\vec p_2$,  which  results in  $f_{\rm eff}\gg \Mp$ although $f_i$ are all sub-Planckian. Note that our parametrization of the inflaton direction can receive a  correction of ${\cal O}(f_i/f_{\rm eff})$, which would give rise to a correction of ${\cal O}(f_i)$ to $f_{\rm eff}$.

Obviously the convex hull condition~\eqref{CHC} for $\vec u=\vec\xi$ cannot be compatible with $f_{\rm eff}\, \gg \, \Mp$ in  \eqref{KNP}, if  the axion-instanton couplings $\{\vec q_I\}$ required by the WGC coincide with the couplings $\{\vec p_I\}$ generating the KNP-type axion potential. Therefore, in order for the alignment mechanism  to be compatible with the WGC, some  of $\{\vec q_I\}$ should not be in $\{\vec p_I\}$~\cite{Brown:2015iha}. In view of that at least some part of the KNP-type axion potential can be induced by either perturbative effects, e.g. flux,  or  nonperturbative effects other than instantons, e.g. hidden quark or gaugino condensations, there is no apparent obstacle to satisfying this condition.  Let $\vec q$ denote such axion-instanton coupling in $\{\vec q_I\}$, which does not participate in the alignment mechanism.  Yet, generically $\vec q$  can induce an additional piece of axion potential
\bea
\Delta V \,=\, \Lambda_{\rm mod}^{4}\left[1-\cos\left(\vec q \cdot \vec \phi+\delta\right)\right]
\label{mod-pot}
\eea
with $\Lambda_{\rm mod}^{4}\ll \Lambda_{1,2}^4$  as this additional potential should not spoil the alignment mechanism. Now the WGC requires that the convex hull spanned by $\{\vec p_I, \vec q\}$ should contain the unit ball. This means that once $\{\vec p_I\}$ are aligned to yield $1/f_{\rm eff}=|\vec \xi\cdot \vec p_1|=|\vec \xi\cdot \vec p_2|\ll 1/\Mp$, there should exist a $\vec q$ satisfying 
\bea
|\vec\xi \cdot \vec q|\, >\, \frac{S_{\rm ins}}{\Mp}.
\label{bound-q}\eea
Again, one can integrate out the heavy axion to derive the effective potential of the light inflaton. With $\vec \phi \,=\,  \phi_{\rm inf}\, \vec\xi$, one finds  that the total inflaton potential is given by
\bea 
V\,=\, V_0+\Delta V \,=\, \Lambda_{\rm eff}^4\left[1-\cos \left(\frac{\phi_{\rm inf}}{f_{\rm eff}}\right)\right]+
\Lambda_{\rm mod}^{4} \left[1-\cos \left(\frac{\phi_{\rm inf}}{f_{\rm mod}}+\delta\right)\right],
\label{inf-mod}
\eea
where \bea
\Lambda_{\rm eff}^4&=&\Lambda_1^4+\Lambda_2^4,\nonumber \\
f_{\rm eff}&=&\frac{1}{|\vec \xi \cdot \vec p_1|}\,=\, \frac{1}{|\vec \xi \cdot \vec p_2|} \,\gg\, \Mp, \nonumber \\
f_{\rm mod}&=& \frac{1}{|\vec\xi\cdot\vec q|}\, <\, \frac{\Mp}{S_{\rm ins}}.\nonumber\eea

\begin{figure}
\centering
\includegraphics[scale=0.48]{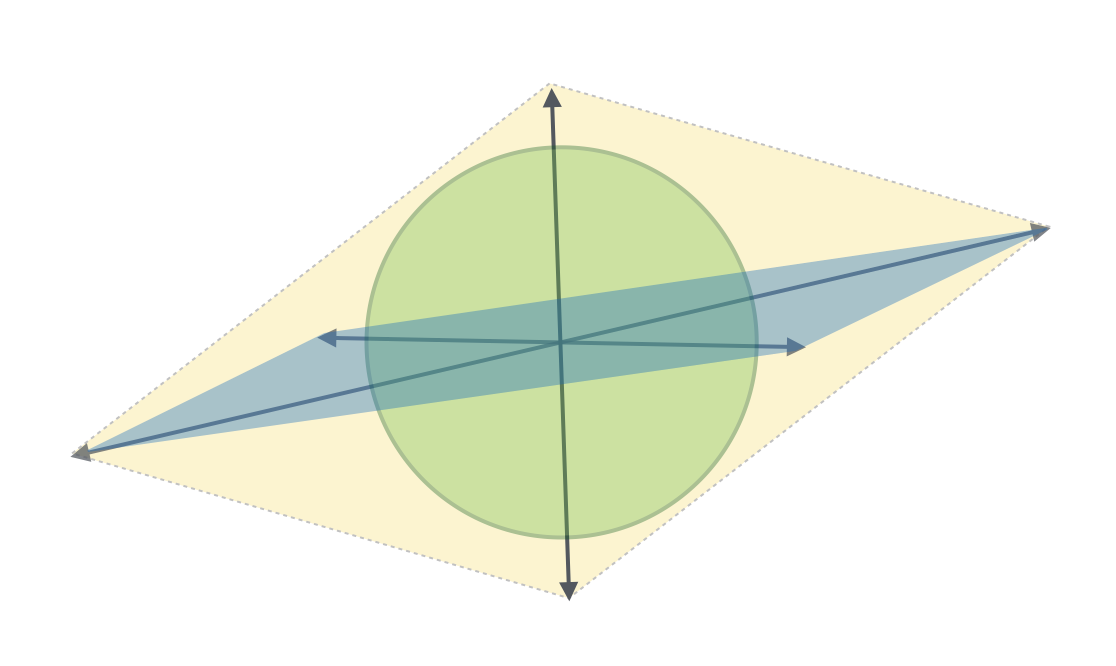}
\caption{The KNP alignment mechanism and the convex hull condition. The axion couplings $\vec{p}_1$ and $\vec{p}_2$ (blue arrows) are aligned to be nearly parallel to produce a super-Planckian effective decay constant. These couplings do not necessarily coincide with the instanton couplings $\vec q_1$ and $\vec q_2$ required by the weak gravity conjecture. Here we assume $\vec q_1=\vec p_1$, and therefore $\vec q=\vec q_2$ (black arrow).}
\end{figure}

The above consideration suggests that small modulation of the inflaton potential is a generic feature of the aligned natural inflation compatible with the WGC. As a specific example for the aligned natural inflation compatible with the WGC, one may consider the axion couplings
 \bea
 \vec{p}_1 =  \left( \frac{\tilde n}{f_1},\, \frac{1}{f_2} \right),\quad
\vec{p}_2 =  \left( \frac{1}{f_1},\, 0 \right), \quad \vec q= \left(0, \frac{1}{f_2}\right)
\eea
with 
$$f_1\sim f_2,\quad \tilde n\gg 1,$$
which corresponds to the axion couplings considered in  \cite{Hebecker:2015rya} to generate a KNP-type axion potential. One then finds
\bea 
\vec \xi = \left( \frac{f_1}{(\tilde{n}-1)f_2}, \, -1 \right)+ \, {\cal O}\left(\frac{1}{\tilde n^2}\right),\quad
f_{\rm eff} = \tilde n f_2 +{\cal O}(f_i),\quad
f_{\rm mod}= f_2 +{\cal O}\left(\frac{f_i}{\tilde n^2}\right).
\eea

It is straightforward to generalize our argument to the case that more axions are involved in the alignment mechanism.  As was stressed in \cite{Choi:2014rja}, in the limit that $N\gg 1$, there can be a more variety of ways to enhance the effective axion decay constant.  For instance,  one can even achieve  $f_{\rm eff}/f_i ={\cal O}( e^N)$, which would make it possible to get a super-Planckian effective decay constant without introducing an unreasonably large number of fields in the model\footnote{Recently this form of exponential hierarchy between the axion scales has been applied for the model of relaxion \cite{Graham:2015cka,Choi:2015fiu,Kaplan:2015fuy}.}. To implement the alignment mechanism with  $\vec\phi=(\phi_1, \phi_2, ...,\phi_N)$, one can consider a potential of the form 
\bea
V_0 =  \sum_{I=1}^N \Lambda_I^4 \left[ 1 - \cos \left( \vec p_I\cdot \vec \phi\right)\right],
\eea
where 
\bea
\vec{p}_I\,=\, \left(\frac{\tilde n_{I1}}{f_1},\frac{\tilde n_{I2}}{f_2}, ..., \frac{\tilde n_{IN}}{f_N}\right)\eea
are linearly independent from each other, but they are aligned in such a way that all $\vec p_I$ lie nearly on an $(N-1)$-dimensional hyperplane. Then the potential has a flat direction which can be parametrized by the inflaton field
\bea
{\phi}_{\rm inf} = \frac{1}{\big(\sum_{i=1}^N a_i^2 \big)^{1/2}  } \, \det \left(
\begin{array}{c}
\vec{\phi} \\
\vec{p}_1- \vec{p}_2 \\
\vdots \\
\vec{p}_{N-1} - \vec{p}_N
\end{array}
\right)\,\equiv \, \vec\xi\cdot\vec \phi,
\eea
where the flat direction unit vector is given by
$$\xi_i =\frac{a_i}{\big(\sum_{i=1}^N a_i^2 \big)^{1/2}}$$
with
\bea
a_i = \det \left(
\begin{array}{ccccccc}
\frac{\tilde n_{11}}{f_1} & \cdots &\frac{\tilde n_{1(i-1)}}{f_{i-1}} & 1 & \frac{\tilde n_{1(i+1)}}{f_{i+1}} & \cdots &\frac{\tilde n_{1N}}{f_N} \\
\frac{\tilde n_{21}}{f_1} & \cdots &\frac{\tilde n_{2(i-1)}}{f_{i-1}} & 1 & \frac{\tilde n_{2(i+1)}}{f_{i+1}} &  \cdots & \frac{\tilde n_{2N}}{f_N} \\
\vdots & &  \vdots & \vdots &\vdots & & \vdots \\
\frac{\tilde n_{N1}}{f_1} & \cdots &\frac{\tilde n_{N(i-1)}}{f_{i-1}} & 1 & \frac{\tilde n_{N(i+1)}}{f_{i+1}} &  \cdots & \frac{\tilde n_{NN}}{f_N} 
\end{array}
\right).
\nonumber
\eea
The effective decay constant of this inflaton field is given by
\bea
f_{\rm eff} = \frac{1}{| \vec{p}_1 \cdot \vec\xi |} =\frac{1}{| \vec{p}_2 \cdot \vec\xi |}=...= \frac{1}{| \vec{p}_N \cdot \vec\xi |} 
= \frac{ \big(\sum_{i=1}^N a_i^2 \big)^{1/2} } {| \det \left(\vec p_1, \vec p_2, ...,\vec p_N\right)^T |},
\label{feff}
\eea
which can have a super-Planckian value  if $\vec p_I$ are aligned to lie nearly on an $(N-1)$ dimensional hyperplane. Note that $\vec\xi$ is normal to $\vec p_{I+1} -\vec p_{I}$ ($I=1,2,...,N-1$), and therefore $|\vec\xi\cdot p_I|$ have a common value for all $I$. 

Again, in order for $f_{\rm eff} \gg \Mp$ to be compatible with the convex hull condition \eqref{CHC}, the axion-instanton couplings $\{\vec q_I\}$ should not coincide with the couplings $\{\vec p_I\}$ generating the KNP-type axion potential. In other words, some of $\{\vec q_I\}$, which will be denoted as $\vec q$, should not belong to $\{\vec p_I\}$. Generically  such axion-instanton coupling  can induce a subleading piece of axion potential, taking the form of   \eqref{mod-pot}. Also, to be compatible with the WGC, the convex hull spanned by $\{\vec p_I,\vec q\}$ should contain the unit ball. This means that for $\{\vec p_I\}$ aligned to generate $f_{\rm eff}\gg \Mp$, there should exist an axion-instanton coupling $\vec q$ satisfying the bound \eqref{bound-q}. Then, after integrating out the $(N-1)$ heavy axions, the resulting inflaton potential includes a modulation part as  \eqref{inf-mod}. Depending upon how the KNP-type axion potential $V_0$ is generated, i.e. depending upon the origin of the axion couplings $\{\vec p_I\}$, there can be multiple  $\vec q$ satisfying the bound \eqref{bound-q}, which would result in multiple modulation terms
\bea
\Delta V \,=\,\sum _{\vec q} \Lambda_{q}^4 \left[1-\cos\left(\vec\xi\cdot \vec q \phi_{\rm inf}+\delta_q\right)\right]\,=\, \sum _{\vec q} \Lambda_{q}^4 \left[1-\cos\left(k_q\frac{\phi_{\rm inf}}{f_{\rm mod}}+\delta_q\right)\right],\eea
where $k_q$ are integers of order unity, and $f_{\rm mod} < \Mp/S_{\rm ins}$.

A particularly  interesting example of the alignment involving $N$ axions has been proposed in \cite{Choi:2014rja}, in which the axion couplings
are given by
\bea \left(
\begin{array}{c}
\vec p_1 \\
\vec{p}_2 \\
\vdots \\ \vec{p}_N
\end{array}
\right) =\left(
\begin{array}{ccccccc}
\frac{1}{f_1} & -\frac{n}{f_2} &    &  &  &  &    \\
   & \frac{1}{f_2} & -\frac{n}{f_3} &  &  &  &    \\
   &    &    & \ddots & & & \\
   &    &    & & & \frac{1}{f_{N-1}} & -\frac{n}{f_N} \\
   &    &    & & &    & \frac{1}{f_N}
\end{array}
\right),\quad \vec q = \left(\frac{1}{f_1}, 0, ...,0\right).
\eea
This results in the flat direction
\bea 
\vec \xi \,\propto\, \frac{n}{n-1} \left( 1-n^{-N} ,\, n^{-1}-n^{-N} ,\, n^{-2}-n^{-N} ,\,  \cdots,\, n^{-(N-1)}-n^{-N}  \right), \eea 
with the effective decay constants 
\bea
f_{\rm eff} \,\sim \, n^{N-1} f_i, \quad f_{\rm mod}\,\sim\,  f_i,
\eea
where all $f_i$ are assumed to be comparable to each other.

\section{Oscillations in primordial power spectrum}\label{cmb}

Our discussion in the previous section suggests that small modulation of the inflaton potential is a generic feature of the aligned natural inflation compatible with the WGC.  In this section, we examine the observable consequence of this modulation and the resulting constraints on the model.

To proceed, we first identify the parameter region of our interest for the inflaton potential with modulation: 
\bea 
\label{inf_pot}
V\,=\, V_0+\Delta V \,=\, \Lambda_{\rm eff}^4\left[1-\cos \left(\frac{\phi}{f_{\rm eff}}\right)\right]+
\Lambda_{\rm mod}^{4} \left[1-\cos \left(\frac{\phi}{f_{\rm mod}}+\delta\right)\right].
\eea
To avoid a fine tuning of the initial condition worse than  10\%, while producing a  spectral index of the CMB power spectrum consistent with the observation, we limit the discussion to  $f_{\rm eff}\gtrsim 5 \, \Mp$. As for the modulation periodicity $f_{\rm mod}$, the WGC suggests that $f_{\rm mod}\lesssim \Mp/S_{\rm ins}$. Quite often, one finds the instanton amplitude $\propto e^{-2\pi T}$,  where $2\pi T = S_{\rm ins}+i\theta$ for an angular axion field $\theta$, and the underlying UV theory reveals the strong-weak coupling duality under $T\rightarrow 1/T$, as well as the discrete shift symmetry: $\theta \rightarrow \theta +2\pi$ \cite{Polchinski:1998rq}.   One then finds $S_{\rm ins}\geq 2\pi$, where the bound is saturated when the associated coupling $\propto 1/S_{\rm ins}$ has a self-dual value with respect to the presumed strong-weak coupling duality. Motivated by these observations, in this section we will concentrate on the parameter region with
\bea
f_{\rm eff}\, \gtrsim \, 5 \Mp, \quad f_{\rm mod} \,\lesssim\, \frac{\Mp}{2\pi}.
\label{f-range}\eea

With the inflaton potential (\ref{inf_pot}), the equation of motion is given by
\bea
\ddot{\phi} + 3 H \dot{\phi} +V_0' (\phi) \left[ 1 + \frac{\Lambda_{\rm mod}^4}{V_0'(\phi)f_{\rm mod}} \sin \left(\frac{\phi}{f_{\rm mod}} + \delta \right) \right] =0.
\eea
The effect of modulation can be treated perturbatively if $\Lambda_{\rm mod}^4 / V_0'(\phi) f_{\rm mod} \ll 1 $. Since we are interested in the inflaton dynamics around when the CMB pivot scale $k_*$ exits the horizon, we define our expansion parameter as~\cite{Flauger:2010ja}
\bea
b \equiv  \frac{\Lambda_{\rm mod}^4}{V_0'(\phi_*) f_{\rm mod}}\,\ll \, 1,
\eea
where $\phi_*$ is the inflaton value when the CMB scale $k_*$ exits the horizon. We then expand the solution of the inflaton field $\phi$ and the corresponding slow-roll parameters as 
\bea
\phi \,=\, \phi_0 + \Delta\phi, \quad
\epsilon \,=\, \epsilon_0 + \Delta\epsilon, \quad
\eta \,=\,  \eta_0 + \Delta\eta,
\eea
where $$\epsilon = -\frac{\dot{H}}{H^2}, \quad  \eta = \frac{\dot{\epsilon}}{H\epsilon},$$ and the modulation-induced corrections $\Delta\phi,\, \Delta\epsilon,\, \Delta\eta$ include only the effects first order in $b$.

To examine the effects of modulation on the curvature perturbation ${\cal R}$, we choose the comoving gauge  for which 
\bea
\delta g_{ij} = 2a^2 {\cal R} \delta_{ij},
\eea
where $a$ and $\delta g_{ij}$ denote the scale factor and the perturbation of spatial metric, respectively. The corresponding evolution equation of the curvature perturbation is given by
\bea
{\cal R}_k ''  -\left( \frac{2+2 \epsilon +\eta }{\tau} \right) {\cal R}_k' + k^2 {\cal R}_k = 0,
\eea
where $\tau$ is the conformal time, $d\tau = dt / a(t)$, and the prime denotes the derivative with respect to $\tau$. The curvature perturbation can be expanded as
\bea
{\cal R}_k (\tau)  =  {\cal R}^{(0)}_k (\tau)  + {\cal R}^{(0)}_k (0)  \, g_k (\tau),
\eea
where ${\cal R}^{(0)}(\tau)$ denotes the curvature perturbation in the absence of modulations, and the correction function $g_k(\tau)$ satisfies 
\bea
g_k '' - \frac{2}{\tau} g_k'+ k^2 g_k = k^2 e^{-ik\tau} ( 2\Delta\epsilon + \Delta\eta).
\eea
Note that ${\cal R}^{(0)}_k(0)$ corresponds to the frozen value of ${\cal R}^{(0)}_k(\tau)$ in the superhorizon limit, $-k\tau \rightarrow 0$. One then finds $g_k(\tau)$  leads to an oscillatory behavior of the curvature  power spectrum~\cite{Flauger:2009ab,Flauger:2010ja}, which can be parametrized as
\bea
{\cal P}_{\cal R} (k) = {\cal P}^{(0)}_{\cal R}(k) \left[ 1 + \delta n_s \cos \left( \frac{\phi_k}{f_{\rm mod}} + \beta\right) \right],
\label{OscPS}
\eea
where  ${\cal P}_{\cal R}^{(0)}(k)$ is the power spectrum in the absence of modulations, which is described well by the standard form 
\bea
{\cal P}^{(0)}_{\cal R}(k) = { A_*^{(0)} } \left(\frac{k}{k_*}\right)^{n_s^{(0)} -1}.
\label{nonosiPS}
\eea
Here  $\delta n_s$, $\beta$, $A_*^{(0)}$, and  $n_s^{(0)}$ are all $k$-independent constants, and $\phi_k$ denotes the inflaton field value when the CMB scale $k$  exits the horizon. Then, from
\bea
{\frac{d\phi_k}{d\ln k} = -\frac{\sqrt{2\epsilon}}{1-\epsilon} \, \Mp },\eea 
we find
\bea 
\cos \left( \frac{\phi_k}{2f_{\rm eff}} \right) \, \simeq\, \left(\frac{k}{k_*}\right)^{\frac{\Mp^2}{2f_{\rm eff}^2}}\cos \left( \frac{\phi_*}{2f_{\rm eff}} \right).
\label{phik}
\eea
It is also straightforward to find
\bea
\delta n_s =
\frac{3 b \sqrt{2\pi \gamma \coth \frac{\pi}{2\gamma}}}{\sqrt{(1+ \frac{3}{2} \, \gamma^2 \Mp^2/f_{\rm eff}^2 )^2 + (3 \gamma )^2}},
\label{dns}
\eea
where
\bea\gamma = \frac{f_{\rm eff}f_{\rm mod}}{\Mp^2} \tan \frac{\phi_*}{2f_{\rm eff}}.\eea
Our result agrees  with ref.~\cite{Flauger:2010ja} in the limit $\gamma \ll1$. See refs.~\cite{Flauger:2009ab,Flauger:2010ja} for a more detailed discussion of the oscillation in the power spectrum.

In the conventional slow roll inflation scenario, one usually assumes that the curvature power spectrum  can be  systematically expanded  in powers of $\ln(k/k_*)$ over the available CMB scales, which would result in 
\bea
{\cal P}_{\cal R}(k) = A_* \left({k}/{k_*}\right)^{n_s -1 + \frac{1}{2}\alpha\ln (k/k_*) + \cdots},
\label{StandardTemplate}
\eea
where $n_s$ and $\alpha$ are $k$-independent constants satisfying 
\bea
\alpha\ln(k_{\rm max}/k_{\rm min}) < |n_s-1|\ll 1, \eea
where $k_{\rm min} <k<k_{\rm max}$ represents the range of the observed CMB scales with $\ln(k_{\rm max}/k_{\rm min})\simeq 6-8$, and the ellipsis stands for the higher order terms which are assumed to be negligible. In the presence of modulations, generically the curvature power spectrum cannot be described by the above form, but requires to introduce an oscillatory piece as in \eqref{OscPS}. One may still ask in which limit the oscillatory piece in \eqref{OscPS} can be mimicked by the conventional form (\ref{StandardTemplate}). To examine this question, let us expand $\phi_k$ in \eqref{phik} as
\bea
\phi_k \,=\, \phi_{*}+\left.\frac{d\phi_k}{d\ln k}\right|_{k_*}\ln \left(\frac{k}{k_*}\right)+ \left. \frac{1}{2}\frac{d^2 \phi_k}{d(\ln k)^2} \right|_{k_*} \ln^2\left(\frac{k}{k_*}\right) + \cdots, 
\eea
which is providing a well controlled approximation for $\phi_k$ as it
is essentially an expansion in powers of $\frac{M_{\rm Pl}^2}{f_{\rm eff}^2}\ln(k/k_*)\ll 1$. In order for the oscillatory piece in the power spectrum \eqref{OscPS} to be well mimicked by the conventional form (\ref{StandardTemplate}),  the sinusoidal functions of $(\phi_k-\phi_*)/f_{\rm mod}$ needs to be well approximated by a simple polynomial of $\ln(k/k_*)$ with  a few terms.
Obviously this is possible only when
 \bea
\frac{\phi_{k_{\rm max}}-\phi_{\rm k_{\rm min}}}{f_{\rm mod}} \,\approx \, \frac{\sqrt{2\epsilon_*}\Mp}{f_{\rm mod}}\ln(k_{\rm max}/k_{\rm min}) \, <\, 1,
\label{condition}\eea
where $\phi_{k_{\rm max}}-\phi_{\rm k_{\rm min}}$ corresponds to the total excursion of $\phi$ over the periods when the observable CMB scales exit the horizon.  If the above condition is satisfied, the curvature power spectrum can be expanded as 
\bea
{\cal P}_{\cal R}(k) =  { A_*^{(0)} } \left(\frac{k}{k_*}\right)^{n_s^{(0)} -1}
\left[1 + \delta n_s \left\{ \cos\left(\frac{\phi_*}{f_{\rm mod}}\right)
-\sin\left(\frac{\phi_*}{f_{\rm mod}}\right)\left(\frac{\phi_k - \phi_*}{f_{\rm mod}}\right) \right. \right.
\nonumber \\
\left. \left.  -\frac{1}{2}\cos \left(\frac{\phi_*}{f_{\rm mod}}\right)\left(\frac{\phi_k - \phi_*}{f_{\rm mod}}\right)^2 +{\cal O}\left( \left(\frac{\phi_k - \phi_*}{f_{\rm mod}}\right)^3\right) \right\} \right],
\eea
where we set $\beta=0$ for simplicity. Comparing this with \eqref{StandardTemplate}, we find that the oscillatory piece in \eqref{OscPS} can be mimicked by the conventional form  \eqref{StandardTemplate} with the following matching conditions:
\bea
{A_*} &=& A_*^{(0)} \Big( 1 + \delta n_s \cos \frac{\phi_*}{f_{\rm mod}} \Big),  \nonumber \\
n_s &=& n_s^{(0)} + \delta n_s \frac{\sqrt{2\epsilon_*}\Mp}{f_{\rm mod}} \sin \left(\frac{\phi_*}{f_{\rm mod}}\right)  ,\nonumber \\
\alpha &=& -\delta n_s \frac{\sqrt{2\epsilon_*}\Mp}{f_{\rm mod}}
 \left[ \frac{\sqrt{2\epsilon_*}\Mp}{f_{\rm mod}} \cos\left(\frac{\phi_*}{f_{\rm mod}}\right) - \frac{\eta_*}{2} \sin\left(\frac{\phi_*}{f_{\rm mod}} \right) \right].
\eea
However, for the parameter region \eqref{f-range} of our interest, the condition (\ref{condition}) is badly violated, so we need to use the parametrization \eqref{OscPS} including the oscillatory piece explicitly, rather than using the conventional form \eqref{StandardTemplate}.

\begin{figure}
\includegraphics[scale=0.42]{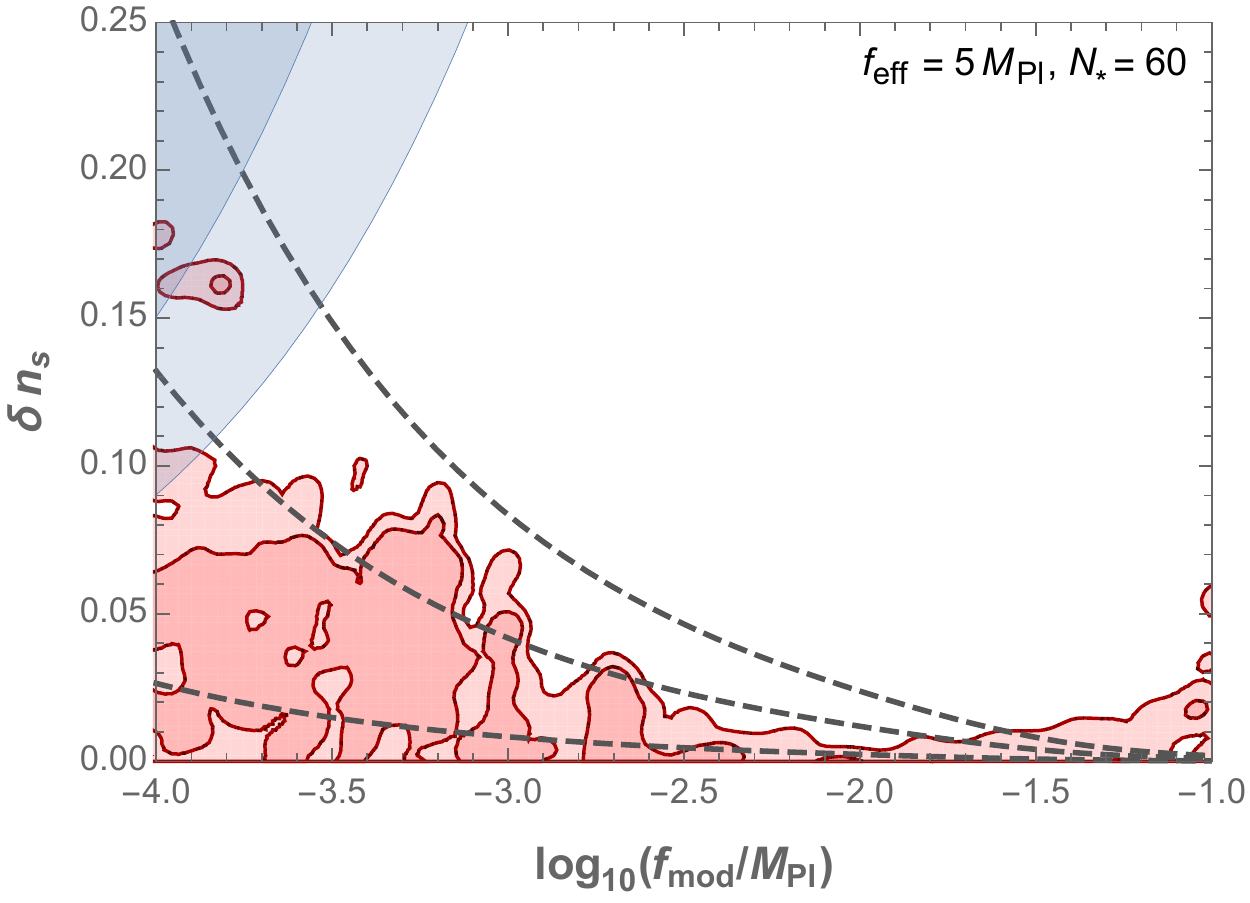}
\includegraphics[scale=0.42]{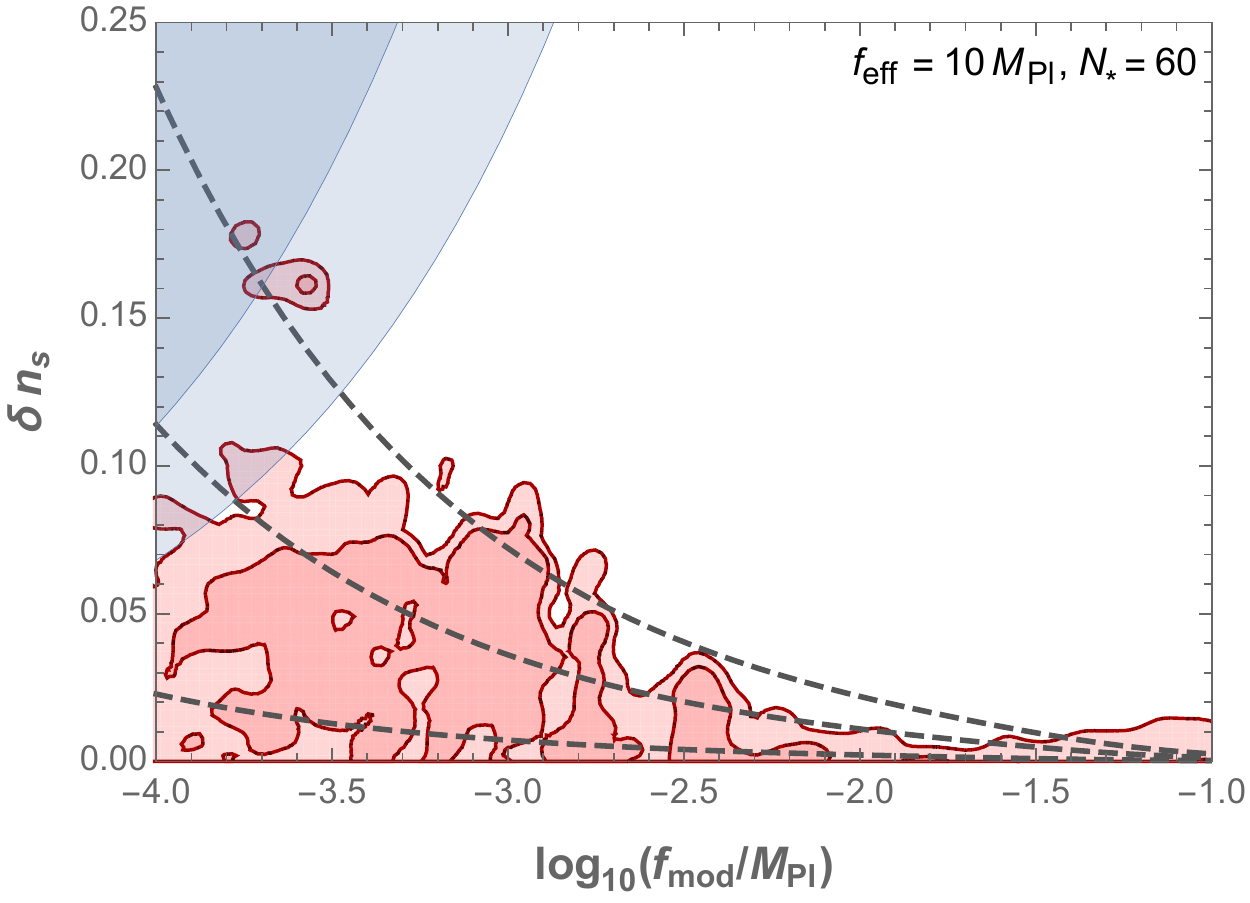}
\includegraphics[scale=0.42]{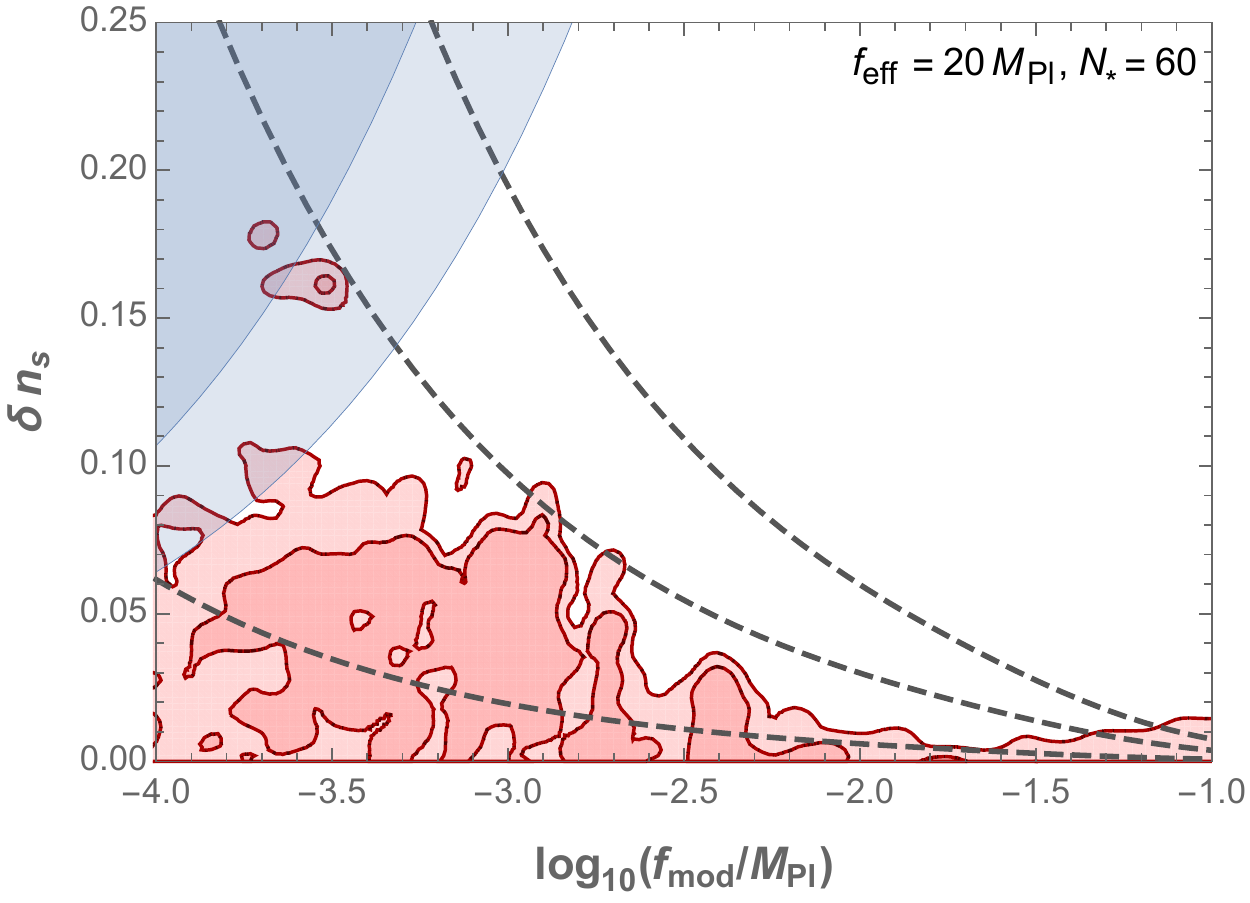}
\caption{Parameter region on the plane of ($f_{\rm mod}, \delta n_s$) with  68\% (pink) and 95\% (light pink) CL likelihood with respect to the Planck data on the temperature anisotropy and low-$\ell$ polarization. The dashed lines represent the predictions from the inflaton potential (\ref{inf_pot})  with $(\Lambda_{\rm mod}/\Lambda_{\rm eff})^4 = 10^{-5},\, 5\times 10^{-6},\, 10^{-6}$ from the top to the bottom. The shaded region in the  upper left corner corresponds to the region that our perturbative approach for modulation becomes  unreliable as the expansion parameter $b$ is not small enough, e.g.  $b \geq 0.3$ for gray region and $b\geq 0.5$ for dark gray region. }
\label{dns_f}
\end{figure}

\begin{figure}
\includegraphics[scale=0.6]{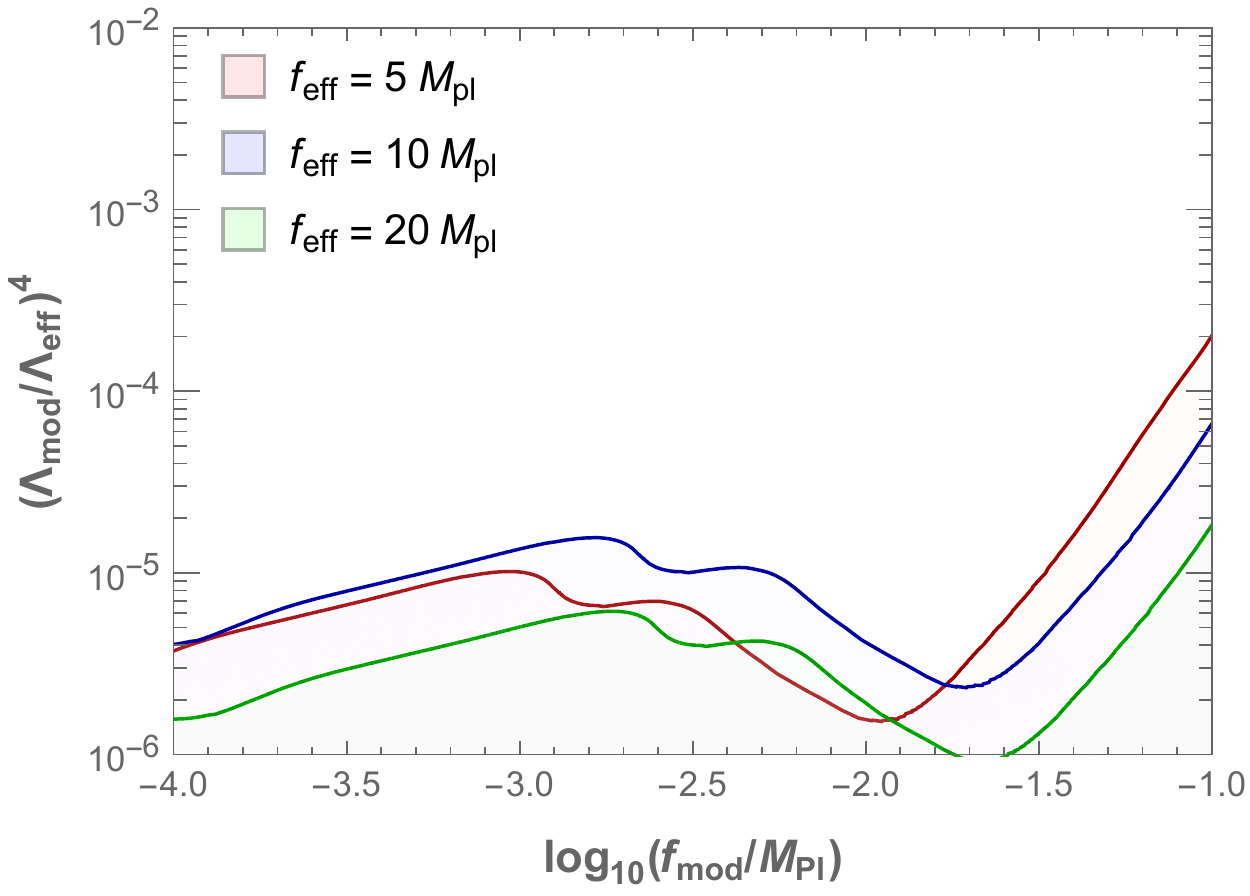}
\caption{$95\%$ CL upper bound on $(\Lambda_{\rm mod}/\Lambda_{\rm eff})^4$ as a function of $f_{\rm mod}$  for $f_{\rm eff} = 5\Mp \textrm{ (red)}$, $\,10\Mp \textrm{ (blue)}$ and $20\Mp$ (green).} 
\label{frac}
\end{figure}

Let us now present the constraints on the model from the CMB data, while taking into account the oscillatory piece in the curvature power spectrum. To this end, we fit the Planck CMB data with the power spectrum~\eqref{OscPS}, and find the likelihood of the phenomenological parameters including  $\delta n_s$ and $f_{\rm mod}$. For our analysis, we use the {\texttt CosmoMC} code~\cite{CosmoMC} with nested sampler {\texttt PolyChord}~\cite{Handley:2015fda}. In Figure~\ref{dns_f}, we show the parameter regions of  $68\%$ and $95\%$ CL in the plane of ($f_{\rm mod},\delta n_s$). Although it depends on the values of $f_{\rm eff}$ and $f_{\rm mod}$, the allowed maximal value of $\delta n_s$ is around $0.1$. The shaded region in the  upper left corner of Figure~\ref{dns_f} corresponds to the region that our perturbative approach for modulation becomes  unreliable as the expansion parameter $b$ is not small enough.  For this region, one needs to compute the primordial power spectrum numerically because the analytic approximation \eqref{OscPS} is not reliable anymore.  

In Figure \ref{frac}, we provide a $95\%$ CL upper bound on $(\Lambda_{\rm mod}/\Lambda_{\rm eff})^4$ as a function of $f_{\rm mod}$  for three different values of  $f_{\rm eff}$\footnote{Here we are focusing on   $5\Mp \lesssim f_{\rm eff} \lesssim 20\Mp$. For  $f_{\rm eff} < 5\Mp$, the model has a difficulty in producing a correct value of the spectral index, and also requires a fine tuning of the initial condition. For $f_{\rm eff} > 20\Mp$, it essentially coincides with the chaotic inflation model with $V_0 =m^2\phi^2$. }: $f_{\rm eff} = 5\Mp , \, 10\Mp,\, 20\Mp$. From this, we find that the amplitude of modulation in the inflaton potential is constrained as $(\Lambda_{\rm mod}/\Lambda_{\rm eff})^4 \lesssim {\cal O}(10^{-4}-10^{-6})$, depending upon the value of the modulation periodicity $f_{\rm mod}$.

\begin{figure}
\includegraphics[scale=0.7]{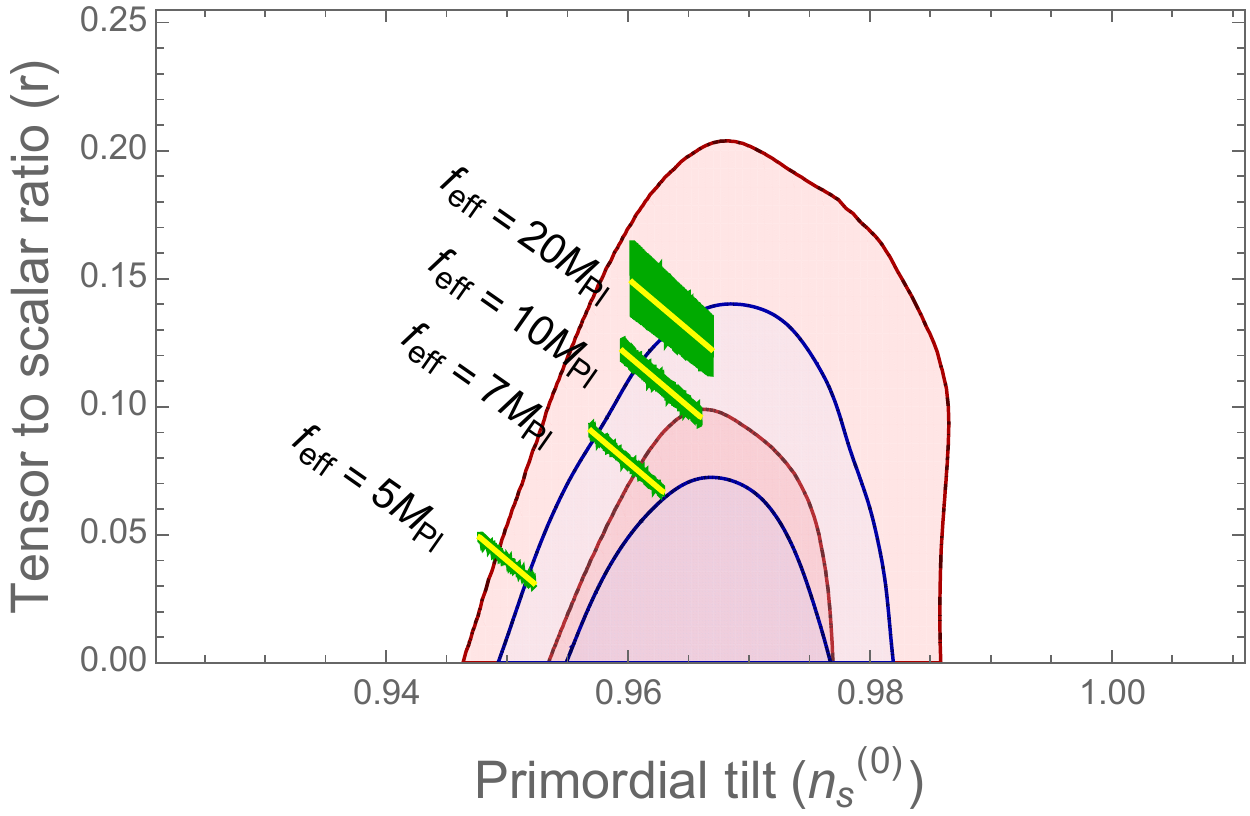}
\caption{
Green bands  represent $(n_s^{(0)}, r)$ predicted by natural inflation with modulations, i.e. the inflaton potential (\ref{inf_pot}), while the yellow lines are the results in the absence of modulation. The red contours represent the model-independent 68\% and 95\% CL ranges of $(n_s^{(0)}, r)$, which are compatible with the observed CMB data fitted with the curvature power spectrum (\ref{OscPS}) including an oscillatory piece. The blue contours are the results of data fitting in the absence of oscillation, i.e. $\delta n_s=0$. }
\label{rFigure}
\end{figure}

As was pointed out in \cite{Abe:2014xja}, modulations may modify the tensor-to-scalar ratio. In the presence of modulation, the scalar power spectrum is modified by $\delta n_s$, while the tensor power spectrum remains almost unaffected. Therefore, in our case, the tensor-to-scalar ratio at the scale $k$ is given by
\bea
r \,= \,\frac{ {\cal P}_t(k)}{ {\cal P}_{\cal R}(k)} \,\simeq \,\frac{16\epsilon_0}{1 + \delta n_s \cos \left(\frac{\phi_k}{f_{\rm mod}} + \beta\right) },
\eea
where $\epsilon_0$ corresponds to the slow-roll parameter in the absence of modulation. Considering the tensor-to-scalar at the pivot scale $k_*$, it can be either enhanced or suppressed because of the oscillating behavior of the scalar power spectrum. In Figure.~\ref{rFigure}, we present the tensor-to-scalar ratio predicted by the aligned natural inflation with modulations, where the spectral index $n_s^{(0)}$ is defined in (\ref{nonosiPS}) for the curvature power spectrum (\ref{OscPS}) involving an oscillatory piece.  The green bands show the results when a nonzero $\delta n_s$, which is given by (\ref{dns}),
   is taken into account.  To be specific, here we choose $(\Lambda_{\rm mod}/\Lambda_{\rm eff})^4 = 5\times 10^{-6}$ and $f_{\rm mod}=10^{-3}\Mp$, however the result is not sensitive to these parameters as long as they are within the observational bound depicted in Figure \ref{frac}.
For comparison, we provide also the results (yellow lines) in the absence of modulation, i.e. $\delta n_s=0$.  Because the oscillation amplitude $\delta n_s$ is limited to be less than about $0.1$, the change of $r$ due to modulations is rather minor, i.e. at most a change of ${\cal O}(10)\%$ compared to the case without modulation.

Although  the  predicted value of $r$ is not significantly affected by modulations, its compatibility with the observed CMB data might be altered. In Figure.~\ref{rFigure}, we depict also the contours representing the 68\% and 95\% CL range of $(n_s^{(0)}, r)$ compatible with the observed CMB data. Red contours are the results when  $\delta n_s$ is allowed to freely vary within the corresponding observational bounds,  while the blue contours are the results when  $\delta n_s$ is frozen to be vanishing. Our results imply that the parameter region of $(n_s^{(0)}, r)$ compatible with the CMB data is enlarged if there were a proper amount of oscillatory piece in the curvature power spectrum. This can ameliorate the potential tension between the natural inflation scenario and the CMB data. For instance, the point with $r=0.05$ and $n_s^{(0)}=0.95$,  which is predicted by natural inflation with $f_{\rm eff}=5 \, M_{\rm Pl}$, is inside the $95\%$ CL contour in the presence of modulation, while it is outside the $95\%$ CL contour in the absence of modulation.

We finally comment on the primordial non-Gaussianity. Modulations of the inflaton potential can give rise to a distinct shape of non-Gaussianity, which is known  as the {\it resonant} non-Gaussianity~\cite{Flauger:2010ja}. In our case, its amplitude is given by
\bea
f^{\rm res}_{NL} = \frac{3b \sqrt{2\pi}}{8 \gamma^{3/2}},
\eea
where $\gamma = (f_{\rm eff} f_{\rm mod}/\Mp^2) \tan (\phi_*/2f_{\rm eff})<1$. This type of non-Gaussianity rarely overlaps with the other types of non-Gaussianity, such as the local, equilateral, and orthogonal non-Gaussianities~\cite{Flauger:2010ja}. By this reason, we cannot simply apply the constraints on the other types of non-Gaussianity for this model. On the other hand, the Planck collaboration has provided an observational constraint on the resonant non-Gaussianity, roughly $f_{NL}^{\rm res} \lesssim 100$ for a narrow range of frequency  with $\gamma \sim 0.1$~\cite{Ade:2013ydc}. In our case, we find $f^{\rm res}_{NL} < {\cal O}(1)$ when $\gamma\sim 0.1$ and the amplitude of modulation satisfies the bound from CMB. Therefore the aligned natural inflation with modulations satisfies easily the constraints on the primordial non-Gaussianities.

\section{conclusion}\label{conclusion}

In this paper, we have provided an argument implying that a small modulation of the inflaton potential  is a generic  feature of the aligned natural inflation consistent with the weak gravity conjecture. We studied also the observable  consequences of modulation in the aligned natural inflation scenario for the theoretically or phenomenologically favored parameter region with $f_{\rm eff} \gtrsim 5 \Mp$ and $f_{\rm mod} \lesssim \frac{1}{2\pi}{\Mp}$.  We find that the Planck CMB data provides a severe bound on the amplitude of modualtion. Although modulation does not cause an appreciable change of the tensor-to-scalar ratio $r$ predicted by the model, it affects the compatibility between the CMB data and the predicted value of $r$, and therefore  can ameliorate the potential tension between the CMB data and the natural inflation scenario.

{\vspace{1cm}
\noindent {\bf Note added:} \\
While this paper was being finalized, we received ref.~\cite{Kappl:2015esy} discussing the aligned natural inflation with modulations within the framework of string theory embedding. Ref.~\cite{Kappl:2015esy} discusses  also the observable consequences of modulations for  lower modulation frequencies than ours. 
}

\section*{Acknowledgement}
We thank Chang Sub Shin for initial collaboration and helpful comments. We  thank also Jinn-Ouk Gong, Raphael Flauger, Seokhoon Yun and Toyokazu Sekiguchi for useful discussions. H.K. thanks Won Sang Cho for his help in running the code. This work was supported by IBS under the project code, IBS-R018-D1.

\end{document}